\documentclass[12pt,aps,prb,floats,eqsecnum,amsmath,graphicx,euscript,amsmath,amssymb,onecolumn] {revtex4}
\usepackage[cp1251]{inputenc}
\usepackage[dvips]{graphicx}
\DeclareMathAlphabet\EuScript{U}{eus}{m}{n} \SetMathAlphabet\EuScript{bold}{U}{eus}{b}{n}

\def\lapprox{\,\raise0.4ex\hbox{$<$}\kern-0.8em\lower0.7ex\hbox{$\sim$}\,}
\def\gapprox{\,\raise0.4ex\hbox{$>$}\kern-0.8em\lower0.7ex\hbox{$\sim$}\,}
\begin{document}
\bibliographystyle{prsty}

\title{Antiphased Cyclotron-Magnetoplasma Mode

in a Quantum Hall System}

\author{L.V.~Kulik$^{1}$, S.~Dickmann$^{1}$, I.K.~Drozdov$^{1}$, I.S.~Zhuravlev$^{1}$,
V.E.~Kirpichev$^{1}$, I.V.~Kukushkin$^{1,2}$, S.~Schmult$^{2}$, and W.~Dietsche$^{2}$}
\affiliation{$^{1}$Institute of Solid State Physics, RAS, Chernogolovka, 142432 Russia\\
$^{2}$Max-Planck-Institut f\"ur Festk\"orperforschung, Heisenbergstr. 1, 70569 Stuttgart,
Germany}

\date{\today}

\begin{abstract}
An antiphased magnetoplasma (MP) mode in a two-dimensional electron gas (2DEG) has been studied
by means of inelastic light scattering (ILS) spectroscopy. Unlike the cophased MP mode it is
purely quantum excitation which has no classic plasma analogue. It is found that zero momentum
degeneracy for the antiphased and cophased modes predicted by the first-order perturbation
approach in terms of the {\it e-e} interaction is lifted. The zero momentum energy gap is
determined by a negative correlation shift of the antiphased mode. This shift, observed
experimentally and calculated theoretically within the second-order perturbation approach, is
proportional to the effective Rydberg constant in a semiconductor material.

\end{abstract}
\maketitle

The unique symmetry properties of the quantum Hall (QH) electron liquid have stimulated progress
in the study of strongly correlated electron systems in perpendicular magnetic field. In
particular, it has been discovered that the simplest excitations of a 2DEG are excitons
consisting of an electron promoted from a filled Landau level (LL) and bound to an effective
hole left in the ``initial'' LL.\cite{by81,by83,ka84} Within the exciton paradigm, the physics
of this many-particle quantum system is reduced to a {\it two-particle} problem. This can be
solved in an asymptotically exact way where the parameter $r_{\rm c}=E_{\rm C}/\hbar\omega_c$ is
considered to be small. Here $E_{\rm C}=\alpha e^2/\kappa l_B$ is the characteristic Coulomb
energy, $\omega_{c}$ is the cyclotron frequency, and the numerical coefficient $\alpha<1$
represents the averaged renormalization factor due to the finite thickness of the 2DEG in
experimentally accessible systems. The excitation energy in this approach is the sum of two
terms: (i) a single-electron gap (which is the Zeeman or cyclotron, or combined one); and (ii) a
correlation shift induced by the electron-electron ({\it e-e}) interaction. Kohn's renowned
theorem dictates that in a translationally invariant electron system one of the excitons
[magnetoplasma (MP) mode] has no correlation shift at $q=0$. This mode is described by the
action of Kohn's ``raising'' operator ${\hat K}_s^\dag=\sum_{np\sigma}\!{}\!\sqrt{n+1}
c^{\dag}_{n\!+\!1,p,\sigma}c_{n,p,\sigma}$ on the 2DEG ground state $|0\rangle$, where
$c_{n,p,\sigma}$ is the Fermi annihilation operator corresponding to the state $(n,p)$ with the
spin index $\sigma\!\!=\!\uparrow,\!\downarrow$ ($n$ is the LL number; $p$ labels the inner LL
number, if, e.g., the Landau gauge is chosen).\cite{ko61} Yet, Kohn's theorem {\it does not ban}
the existence of another homogeneous MP mode that has a non-vanishing correlation shift.
Precisely two MP modes should coexist at odd electron fillings $\nu\!>\!1$ when the numbers of
fully filled spin sublevels differ by unit, see the illustration in Fig.~1. The symmetric mode
is a {\it cophased} (CP) oscillation of spin-up and spin-down electrons, and the anti-symmetric
one is an {\it antiphased} (AP) oscillation of two spin subsystems. When calculated to first
order in terms of the parameter $r_{\rm c}$, Kohn's mode (the CP magnetoplasmon) has the
energy$\,$\cite{ka84,ch74}
$$
  E_s(q)\!=\!\hbar\omega_c\!+\!\nu e^2q/2\kappa\!+\! O(E_{\rm C}q^2l_B^2)  \eqno (1)
$$
at small $q$ ($ql_B\!\ll\! 1$). The AP mode is a state orthogonal to ${\hat K}_s^\dag|0\rangle$.
It has the energy $E_a(q)=\hbar\omega_c\!+\!O(E_{\rm C}q^2l_B^2)$ calculated to first order in
$r_{\rm c}$.\cite{ka84} Both Coulomb shifts, $\Delta_{s,\,a}=E_{s,\,a}(0)-\hbar\omega_c$, thus
vanish if calculated up to $\sim r_{\rm c}$. So, within this approximation, both MP modes turn
out to be degenerate at $q=0$.

Kohn's MP mode has been a prime subject for the cyclotron resonance studies, and the validity of
Kohn's theorem has been confirmed scores of times.\cite{an82} It is well established
experimentally that homogenous electromagnetic radiation incident on a translationally invariant
electron system is unable to excite internal degrees of freedom associated with the Coulomb
interaction, i.e. $\Delta_s\equiv0$. No similar experiments have been performed for the AP mode
as it is not active in the absorption of electromagnetic radiation. Recent development of
Raman scattering spectroscopy to the point when it became sensitive to the cyclotron spin-flip
and spin-density excitations$\,$\cite{er99,ku05,va06} opened the opportunity to employ this spectroscopy in the
investigation of the AP mode. Here, we report on a direct observation of the AP mode for a
number of odd electron fillings and show that the theoretically predicted zero momentum degeneracy
for Kohn's and AP modes is in fact lifted due to many particle correlations. We also show that the {\it second-order
corrections} to the excitation energies accurately reproduce the observed effect. The
correlation shift for the AP mode is non-vanishing and negative at $q=0$.\cite{foot2}

Several high quality heterostructures were studied. Each consisted of a narrow $18\div 20$~nm
GaAs/Al$_{0.3}$Ga$_{0.7}$As quantum well (QW) with an electron density of $1.2\div 2.4 \times
10^{11}$\,cm$^{-2}$. The mobilities were $3\div 5 \times 10^{6}$\,cm$^2$/V$\cdot$s -
very high for such narrow QWs. The electron densities were tuned via the opto-depletion
effect and were measured by means of in-situ photoluminescence. The experiment was performed at a
temperature of $0.3$\,K. The QWs were set on a rotating sample holder in a cryostat with
a 15\,T magnet. The angle between the sample surface and the magnetic field was
varied in-situ. By continuously tuning the angle we were able to increase the Zeeman energy
while keeping the cyclotron energy fixed. This reduced thermal spin-flip excitations through the
Zeeman gap. The ILS spectra were obtained using a Ti:sapphire laser tunable above the
fundamental band gap of the QW. The power density was below $0.02$\,W/cm$^2$. A two-fiber
optical system was employed in the experiments.\cite{ku06} One fiber transmitted the pumping
laser beam to the sample, the second collected the scattered light and guided it out of
the cryostat. The scattered light was dispersed by a Raman spectrograph and recorded with a
charge-coupled device camera. Spectral resolution of the system was about 0.03\,meV.

Narrow QWs were chosen to maximize energy gaps separating the size-quantized electron subbands.
This mitigated the subband mixing induced by the tilted magnetic field. Yet, the mixing effect
was important and we put it under close scrutiny. The influence of the tilted magnetic field on
the cyclotron energy was studied for every QW by measuring the energies and dispersions for the MP
and Bernstein modes.\cite{ku06} Most accurately this procedure was performed for the narrowest $18\,$nm QW
where the non-linearity was fairly small. Besides, it is exactly the $18\,$nm QW where
the correlation shift reaches its largest value, as it is affected by the QW width through
the renormalization factor $\alpha$. Therefore, hereafter we will only address the $18\,$nm QW.

The ILS resonances for both CP and AP modes are shown in Fig.~1. They have quite different
properties. Kohn's resonance is blue shifted from the cyclotron energy. Its small momenta
dispersion is given by Eq. (1). Experimentally  $q$ is defined by the orientation of pumping and
collecting fibers relative to the sample surface. It is $0.7\cdot10^5$~cm$^{-1}$ for the spectra
in Fig.~1. Kohn's resonance is well broadened because of linear $q$-dispersion (1), and because
the momentum is effectively integrated in the range of $q\sim 0.6\div 0.8\cdot10^5\,$cm$^{-1}$
due to the finite dimension of the fibers. On the contrary, the resonance for the AP mode is red
shifted and does not broaden. In fact, we did not see any appreciable change in the AP mode
energy upon varying the momentum transferred to the 2DEG via the ILS process.  This experimental
finding agrees with the first order perturbation theory of Ref.\onlinecite{ka84} which predicts
a negligible (compared to the experimental resolution) change of the AP mode energy at small
$q$, defined by the light momentum. Variation of the AP shift in the accessible range of
magnetic fields and electron densities is also within the experimental uncertainty. Since
dimensional analysis of second order Coulomb corrections to the energies of inter-LL excitations
yields exactly an independence of the correlation shift on the magnetic field, we assume that
the origin of the AP shift should be sought within the second order perturbation
theory.\cite{ku05}

\begin{figure}[h]\begin{center} \vspace{-3.mm}
\includegraphics*[width=.6\textwidth]{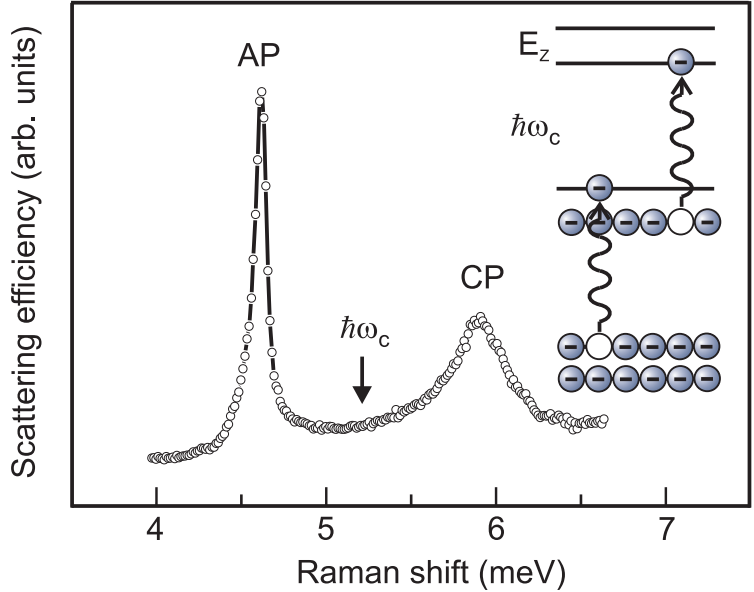}
\end{center}
\vspace{-10.mm} \caption{ILS spectra of Kohn's (CP) and antiphased (AP) magnetoplasma modes
taken at $\nu=3$. The arrow indicates the cyclotron energy. The picture illustrates two single
electron transitions at odd filling factors that, when coupled by the Coulomb interaction, give
rise to two magnetoplasma modes. } \label{f.1} \vspace{-1.5mm}
\end{figure}

The red shift for the AP mode at odd $\nu$ (QH {\it ferromagnets}) is filling factor dependent,
it reduces at larger $\nu$ (Fig.~2). Interestingly its value falls on the same $1/\nu$ curve
that describes the correlation shifts for the antisymmetric mode in another QH system, namely
that for the {\it cyclotron spin-flip mode in a spin-unpolarized 2DEG} at even $\nu$  (Fig.~3).
These two kinds of excitations differ by the total spin quantum number: $S=0$ for the AP mode
which is a spinless magnetoplasmon, and $S=1$ for the cyclotron spin-flip mode. The latter
splits into three Zeeman components with different spin projections along the magnetic field. As
a consequence, in the experimental spectra of Fig.~2 a single ILS resonance corresponds to the
AP mode, whereas the cyclotron spin-flip mode is represented by the Zeeman triplet. The {\it
e-e} correlation nature of red shift for the cyclotron spin-flip mode is confirmed theoretically
in our previous publications,\cite{ku05,di05} and here we employ a similar approach to calculate
the AP shift at $\nu\!=\!3$.

\begin{figure}[h]\begin{center} \vspace{-5.mm}
\includegraphics*[width=.5\textwidth]{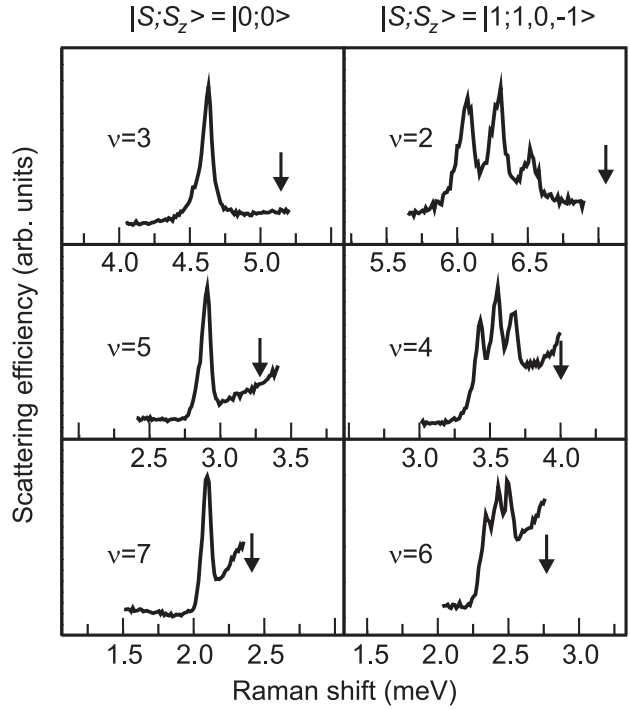}
\end{center}
\vspace{-10.mm} \caption{ILS spectra of the AP magnetoplasma mode (left) and three Zeeman components of the cyclotron
spin-flip mode (right) taken at odd (left) and even (right) filling factors. The arrows
indicate the corresponding cyclotron energies. } \label{f.2} \vspace{-1.5mm}
\end{figure}

Our technique is a variation of the standard perturbative
technique,\cite{ll} although it has some special features. The first is the usage of
the excitonic representation,\cite{di05,di02} where the basis of exciton states is employed
instead of degenerate single-electron LL states. Second, in the development of the perturbative
approach one is forced to use a non-orthogonal basis of two-exciton states. These are
created by action of the interaction Hamiltonian on the single-exciton basis, when considering
first-order corrections to the exciton states. The third feature lies in calculating the exciton
shift counted from the ground state energy, and the latter also has to be taken into account up
to the second order corrections.

Because of the two-fold degeneracy of the $q\!=\!0$ MP states we have to employ two
single-exciton states as a bare basis set. As a result, we come to a $2\!\times \!2$ secular
equation. The bare states are $|X_\downarrow\rangle=Q^\dag_{\overline{0}\overline{1}}|0\rangle$
and $|X_\uparrow\rangle=Q^\dag_{12}|0\rangle$, where $Q_{mk}^\dag=\left.{\EuScript Q}_{mk\,{\bf
q}}^\dag\right|_{{\bf q}=0}$, and $Q_{\overline{mk}}^\dag=\left.{\EuScript
Q}_{\overline{mk}\,{\bf q}}^\dag\right|_{{\bf q}=0}$, and the exciton operators are defined,
e.g., as$\,$\cite{di05,dz83,di02}
$$
  {\EuScript Q}_{mk\,{\bf q}}^{\dag}=\frac{1}{\sqrt{ N_{\phi}}}\sum_{p}\,
  e^{-iq_x\!p}\,
  c_{k,p+\frac{q_y}{2},\uparrow}^{\dag}\,c_{m,p-\frac{q_y}{2},\uparrow}\, \eqno (2)
$$
(${\EuScript Q}_{\overline{mk}\,{\bf q}}^\dag$ differs by changing $\uparrow$ to $\downarrow$ in
the r.h.s.); $q$ is measured in units of $1/l_B$, $N_\phi$ is the LL degeneracy number. The
commutation rules of exciton operators define a special Lie algebra. Considering ${\hat {\cal
H}}_{\rm int}$ as a part of the interaction Hamiltonian relevant to the calculation of the
second-order energy corrections, we present it as a combination of two-exciton operators
$$
{\hat{\cal H}}_{\rm int}=\frac{e^2}{2\kappa l_B}\sum_{n_1\!,\,n_2\!,\,{\bf q}\atop
m_1\!,\,m_2}\left({\hat H}^{\downarrow\downarrow}_{n_1n_2\,{\bf q}\atop
m_1m_2}\!\!\!\!\!\!\!{\vphantom{({\hat \Sigma})}}^\dag\:\:\:+2{\hat
H}^{\downarrow\uparrow}_{n_1n_2\,{\bf q}\atop m_1m_2}\!\!\!\!\!\!\!{\vphantom{({\hat
\Sigma})}}^\dag \:\:\:+{\hat H}^{\uparrow\uparrow}_{n_1n_2\,{\bf q}\atop
m_1m_2}\!\!\!\!\!\!\!{\vphantom{({\hat \Sigma})}}^\dag \:\:\:\right), \eqno (3)
$$
where ${\hat H}^{\downarrow\uparrow}_{n_1n_2\,{\bf q}\atop
m_1m_2}\!\!\!\!\!\!\!{\vphantom{({\hat \Sigma})}}^\dag \:\:\:=V(q)h_{m_1n_1}({\bf
q})h_{m_2n_2}(-{\bf q}){\EuScript Q}^\dag_{\overline{m_1n_1\!{}}\,\,{\bf q}}{\EuScript
Q}^\dag_{{m_2n_2}-{\bf q}}$, $\; 2\pi V({q})$ is the dimensionless 2D Fourier component of the
Coulomb potential, $h_{mn}({\bf q})=({m!}/{n!})^{1/2}e^{-q^2/4}
  (q_-)^{n\!-\!m}L^{n\!-\!m}_{m}(q^2/2)$ [$L_m^n$ is the Laguerre polynomial,
  $q_{\pm}=\mp\frac{i}{\sqrt{2}}(q_x\pm iq_y)$]. Expressions for the first and the third
  operators in parentheses in Eq. (3) differ from the expression for
  ${\hat H}^{\downarrow\uparrow}_{...}{\vphantom{({\hat \Sigma})}}^\dag $
by replacement of  ${\EuScript Q}^\dag$-operators' indexes: $m_2n_2\to\overline{m_2n_2}$, and
$\overline{m_1n_1}\to m_1n_1$ correspondingly. Besides, we may define that  { $ {\hat
H}^{\uparrow\downarrow}_{...}{\vphantom{({\hat \Sigma})}}^\dag\equiv {\hat
H}^{\downarrow\uparrow}_{...}{\vphantom{({\hat \Sigma})}}^\dag$.} As a result of a consistent
perturbative study we find that the correct zero-order MP states
$C_\downarrow|X_\downarrow\rangle+C_\uparrow|X_\uparrow\rangle$ and the correlation shifts are
obtained from the equation
$$
  {\cal E}C_\sigma=\sum_{\sigma'}C_{\sigma'}M_{\sigma\sigma'}, \eqno (4)
$$
where the quantities $M_{\sigma\sigma'}^{(1)}=\langle X_\sigma|{\hat H}_{\rm
int}|X_{\sigma'}\rangle-E_0^{(1)}\delta_{\sigma\!,\,\sigma'}$, calculated within the first-order
approximation, vanish ($E_0^{(1)}$ is the ground state energy calculated to the first order),
whereas the second-order approximation yields \vspace{-5mm}
$$\begin{array}{r}
M_{\downarrow\downarrow}=-\displaystyle{ \frac{({e^2}/{\kappa l_B})^2}{4\hbar\omega_c}
\sum_{\sigma_1\!,\,\sigma_2} \sum_{n_1\!,\,n_2\!,\,{\bf q}\atop m_1\!,\,m_2}
\sum_{n_1'\!,\,n_2'\!,\,{\bf q}'\atop m_1'\!,\,m_2'} \left(\frac{\left\langle
0\left|\left[Q_{\overline{0}\overline{1}}\;,{\hat H}^{\sigma_1\sigma_2}_{n_1'n_2'\,{\bf q}'\atop
m_1'm_2'}\!\!\!\!\!{\vphantom{({\hat \Sigma})}}\:\:\:\right]\left[{\hat
H}^{\sigma_1\sigma_1}_{n_1n_2\,{\bf q}\atop m_1m_2}\!\!\!{\vphantom{({\hat
\Sigma})}}^\dag\:, Q_{\overline{0}\overline{1}}^\dag\right]\right|0\right\rangle}{n_1+n_2-m_1-m_2} \right.}\\
\displaystyle{\left.+N_\phi^{-1/2}\frac{\left\langle 0\left|{\hat
H}^{\sigma_1\sigma_2}_{n_1'n_2'\,{\bf q}'\atop m_1'm_2'}{\vphantom{({\hat
\Sigma})}}\left[Q_{\overline{00}}-Q_{\overline{11}}\,,     {\hat
H}^{\sigma_1\sigma_1}_{n_1n_2\,{\bf q}\atop m_1m_2}\!\!\!{\vphantom{({\hat
\Sigma})}}^\dag\:\right]\right|0\right\rangle}{n_1+n_2-m_1-m_2}\right)\,.}
\end{array} \eqno (5)
$$
The LL number indexes $n_i,\,n'_i$ and $m_i,\,m_i'$ run from 0 to infinity, however only terms
for which $n_1\!+\!n_2\!-\!m_1\!-\!m_2\!=\!n_1'\!+\!n_2'\!-\!m_1'\!-\!m_2'\geq 1$ contribute to
the total sum (5). (Other terms, being not subject to this condition, have zero numerators.) The
expression for another diagonal matrix element $M_{\uparrow\uparrow}$ differs from Eq. (5) by
replacements $\:Q_{\overline{01}}\,\to\, Q_{{12}}$, $\:Q_{\overline{01}}^\dag\,\to\,
Q_{{12}}^\dag$, $\:Q_{\overline{00}}\,\to\,Q_{{11}}$, and $\:Q_{\overline{11}}\,\to\,Q_{{22}}$,
whereas the non-diagonal element $M_{\downarrow\uparrow}$ differs from expression (5) by the
absence of the second term in parentheses and the change from $Q^\dag_{\overline{01}}$ to
$Q^\dag_{12}$ in the first term. Correspondingly, $M_{\uparrow\downarrow}$ is also obtained by
omitting the second term and replacing $Q_{\overline{01}}$ with $Q_{12}$. Analysis shows that
$M_{\downarrow\uparrow}\equiv M_{\uparrow\downarrow}$, as it should be (both values are real).

Fortunately, the symmetry of the system and Kohn's theorem simplify the calculations a great
deal. First, note that one solution of Eqs. (4) is actually known. Indeed, the CP magnetoplasma
mode in the zero order is written as ${\hat K}_s|0\rangle\!\equiv\!\sqrt{N_\phi}\left(
|X_\downarrow\rangle\!+\!\sqrt{2}|X_\uparrow\rangle\right)$. Therefore, substituting
$C_\downarrow=1$, $C_\uparrow=\sqrt{2}$ and ${\cal E}={\Delta}_s\equiv 0$ into Eqs. (4), we
obtain two necessary identities: $M_{\downarrow\uparrow}\equiv
-M_{\downarrow\downarrow}/\sqrt{2}$ and $M_{\uparrow\uparrow}\equiv
M_{\downarrow\downarrow}/2$.\cite{foot3} Another root of the secular equation, ${\rm
det}\!\left|M_{\sigma_1\sigma_2}\!-{\cal E}\delta_{\sigma_1,\,\sigma_2}\right|\!=\!0$, is just
the correlation shift for the AP mode and thus expressed in terms of the only matrix element
(5): ${\cal E}\!=\!\Delta_a=\! 3M_{\downarrow\downarrow}/2$. Second, considerable
simplifications occur in the calculations associated with Eq. (5). It is evident that the ${\hat
H}^{\uparrow\uparrow}_{...}{\vphantom{({\hat \Sigma})}}^\dag $ terms commuting with
$Q$-operators in Eq. (5) do not contribute to the result. However, due to Kohn's theorem, the
${\hat H}^{\downarrow\downarrow}_{...}{\vphantom{({\hat \Sigma})}}^\dag $ operators do not
contribute either. Indeed, consider our ground state as a direct product of two fully polarized
ground states: $|\,0\rangle\!\equiv\!|\,0\!\downarrow\rangle\!\otimes\!|0\!\uparrow\rangle$.
Here $|\,0\!\!\downarrow\rangle$ is the $\nu\!=\!1$ ground state with a positive $g$-factor, and
$|\,0\!\uparrow\rangle$ is the $\nu\!=\!2$ QH ferromagnet  realized in the situation when the
$g$-factor is negative but the Zeeman gap is larger than the cyclotron gap. In Eq. (5) all terms
with the ${\hat H}^{\downarrow\downarrow}_{...}{\vphantom{({\hat \Sigma})}}^\dag $ operators act
only on the $\nu\!=\!1$ ground state and, taken together, {\it yield zero}, because sum of these
terms would constitute the $q=0$ correlation shift of Kohn's mode for the $\nu\!=\!1$ QH
ferromagnet.

Substituting the terms ${\hat H}^{\downarrow\uparrow}_{...}{\vphantom{({\hat \Sigma})}}^\dag $
and ${\hat H}^{\uparrow\downarrow}_{...}{\vphantom{({\hat \Sigma})}}^\dag $ into Eq. (5) and
calculating the commutators according to commutation rules for exciton operators,\cite{di05} one
finds
$$
  {\Delta}_a=-\frac{3m_e^*e^4}{2\kappa^2\hbar^2}\int_0^\infty\!\! qdq V\!(q)^2G(q)\,,  \eqno (6)
$$
where
%\end{document}
$$\begin{array}{r}
  G(q)=\displaystyle{\sum_{n_2=2}^\infty\left[\frac{|h_{1n_2}|^2(h_{00}^2\!-\!2h_{00}h_{11})}{n_2-1}+
\frac{|h_{0n_2}|^2(h_{00}^2-2h_{00}h_{11})\!-\!|h_{01}h_{1n_2}|^2}{n_2}
-\frac{|h_{01}h_{0n_2}|^2}{n_2+1}\right.}\\
\displaystyle{\left.+\sum_{n_1=1}^\infty\left(\frac{|h_{1n_1}h_{1n_2}|^2}
{n_1\!+\!n_2\!-\!2}+\frac{|h_{1n_1}h_{0n_2}|^2\!-\!|h_{0n_1}h_{1n2}|^2}{n_1\!+\!n_2\!-\!1}-
\frac{|h_{0n_1}h_{0n_2}|^2}{n_1\!+\!n_2\!}\right)\right]}\,.
\end{array}  \eqno (7)
$$
We emphasize that this result for $\Delta_a$ includes all contributions to the second-order
correction. In Eq. (7) terms containing only squared moduli of the $h$-functions yield the
direct Coulomb contribution. Terms containing $...h_{00}h_{11}$ are of exchange origin. (Thus
the exchange contribution to the correlation shift is positive.)

In the strict 2D limit, $V(q)\!=\!1/q$, and the correlation shift (6)-(7) is equal to $-0.1044$
if expressed in the 2Ry${}^*\!\!=\!{m_e^*e^4}/{\kappa^2\hbar^2}\!\approx\! 11.34\,$meV units.
This value is nearly $2/3$ of the correlation shift for the $\nu\!=\!2$ cyclotron spin-flip mode
$\Delta_{\mbox{\scriptsize SF}}\!=\!-0.1534$,\cite{di05} which is in surprisingly good agreement
with the experimental $1/\nu$ dependence. Finally, substituting $V(q)=F(q)/q$ into Eq. (6), one
obtains a numerical result for the correlation shift of the zero momentum AP mode at
$\nu\!=\!3$, see Fig.~3. Here, the formfactor $F(q)$ is calculated with the usual
self-consistent procedure\cite{lu93}. The calculation result looks quite satisfactory compared
to the ILS data, if one takes into account that under specific experimental conditions the
quantity $r_{\rm c}$ can only be considered  as a ``small parameter'' with great reserve.

\begin{figure}[h]\begin{center} \vspace{-3.mm}
\includegraphics*[width=.6\textwidth]{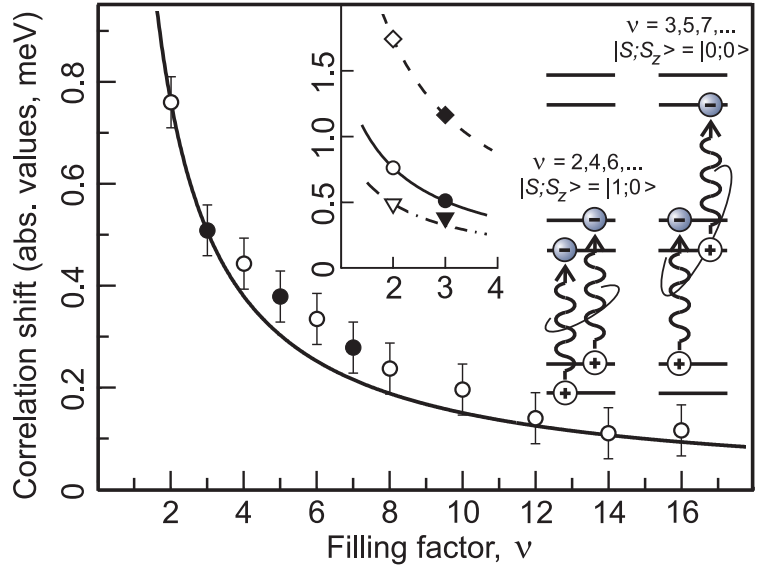}
\end{center}
\vspace{-10.mm} \caption{Main picture: correlation shifts for the AP magnetoplasma mode (solid
dots) and for the $|\,S;S_z\rangle\!=\!|\,1;0\rangle\!$ component of the cyclotron spin-flip
mode (open dots). The solid line shows the $2|\Delta_{\mbox{\scriptsize SF}}^{(\nu=2)}|/\nu$
dependence. In the inset, theoretical values for $|\Delta_{\mbox{\scriptsize SF}}|$ and
$|\Delta_a|$ at $\nu\!=\!2$ and $\nu\!=\!3$ found for the self-consistently computed formfactor
$F(q)$ (triangles) and for the strict 2D limit $F(q)=1$ (diamonds). Corresponding functions
$2|\Delta_{\mbox{\scriptsize SF}}^{(\nu=2)}|/\nu$ are shown by dashed and dot-dashed lines.
Circles and the solid line represent the experimental data. Illustration of single-electron
transitions involved in the $|\,S;S_z\rangle\!=\!|\,1;0\rangle\!$ component of the cyclotron
spin-flip triplet ($\nu=2,4,6,...$) and in the AP mode ($\nu=3,5,7,..$) is given on the right. }
\label{f.3} \vspace{-1.5mm}
\end{figure}

To conclude, we outline the general meaning of the presented results. It is known that optical
methods (including ILS), being in practice the only tool for {\it direct study} of cooperative
excitations in a correlated 2DEG, suffer from an inevitable disadvantage: small momenta of
studied excitations, are far off the interesting region corresponding to inverse values of mean
electron-electron distance. Besides, studying the symmetric MP spectra, one only comes to the
results well described by the classical plasma formula (1), which can be rewritten as
$E_s/\hbar\approx \omega_c\!+\!\Omega_p^2/2\omega_c$ ($\Omega_p$ to denote the 2D plasma
frequency). Therefore the CP magnetoplasma modes are actually classical plasma oscillations
irrelevant to any quantum effects. Contrary to this, homogeneous but antisymmetric modes, namely
the AP mode in a QH ferromagnet and the cyclotron spin-flip mode in an unpolarized QH system are
{\it quantum excitations} even at zero $q$ --- related to the existence of both the spin-up and
spin-down subsystems. The correlation shift, measured in effective Rydbergs, represents
therefore a purely quantum effect. In particular, it includes exchange corrections, which can be
taken into account neither by classical plasma calculations nor by the random phase
approximation (RPA) approach. Quantum origin, common for both types of antisymmetric excitation,
seems to be a reason why both second-order correlation shifts are empirically well described by
the same $1/\nu$ dependence shown in Fig.~3.

The authors thank A.~Pinczuk and A.B. Van'kov for useful discussion and acknowledge support from
the Russian Foundation for Basic Research, CRDF, and DFG.


\begin{references}
\bibitem{foot2} Positive value of ${\Delta}_a$ would not remove the degeneracy but
only shifts the degeneracy point to a non-zero value of $q\sim r_{\rm c}/l_B$. Such a feature
would be physically unjustified.
\bibitem{foot3} Calculation of all elements $M_{\sigma_1\sigma_2}$
can be performed by means of formula (5) and by similar ones. This calculation, giving values
satisfying the necessary identities, serves as a good check for the correctness of our theory.
\end{references}
\end{document}